\documentclass[fleqn,10pt]{wlscirep}

\usepackage[utf8]{inputenc}
\usepackage{amsmath,mathpazo}
\usepackage{graphicx}
\usepackage{textgreek}
\usepackage{verbatim}
\usepackage{hyperref}
\usepackage{enumitem}

\DeclareMathOperator*{\argmax}{arg\,max}
\setlist{nolistsep}

\title{Design of metalloproteins and novel protein folds using variational autoencoders}

\author[1,+]{Joe G Greener}
\author[1,+]{Lewis Moffat}
\author[1,*]{David T Jones}

\affil[1]{Department of Computer Science, University College London, Gower Street, London WC1E 6BT; and Francis Crick Institute, 1 Midland Road, London NW1 1AT}

\affil[+]{these authors contributed equally to this work}

\affil[*]{d.t.jones@ucl.ac.uk}


\begin{abstract}
The design of novel proteins has many applications but remains an attritional process with success in isolated cases.
Meanwhile, deep learning technologies have exploded in popularity in recent years and are increasingly applicable to biology due to the rise in available data.
We attempt to link protein design and deep learning by using variational autoencoders to generate protein sequences conditioned on desired properties.
Potential copper and calcium binding sites are added to non-metal binding proteins without human intervention and compared to a hidden Markov model.
In another use case, a grammar of protein structures is developed and used to produce sequences for a novel protein topology.
One candidate structure is found to be stable by molecular dynamics simulation.
The ability of our model to confine the vast search space of protein sequences and to scale easily has the potential to assist in a variety of protein design tasks.
\end{abstract}

\begin{document}

\flushbottom
\maketitle

\thispagestyle{empty}

\section*{Introduction}

The computational design and redesign of proteins provides a route to create new protein structures and functions \cite{Huang2016, Samish2011}. The `inverse folding problem' of finding a sequence that folds to a given structure or carries out a given function is challenging due to the vast number of possible sequences, the difficulty of assessing the suitability of a sequence, and the marginal stability of protein structures \cite{Yue1992}. From early attempts to add functional motifs to existing structures \cite{Regan1995}, the field has progressed through design of a novel fold \cite{Kuhlman2003} to the use of high-throughput assays to test thousands of sequences simultaneously \cite{Rocklin2017, Chevalier2017}.

Metalloproteins have been shown to be incredibly abundant and important to cell function \cite{Andreini2013}. Designing metalloproteins offers the potential not only to improve our understanding of their function but also to develop new applications within research and industrial settings. Designing metal binding sites in proteins is a relatively mature research area \cite{Regan1995,Andreini2009} and a variety of methods have been succesfully applied ranging from the deterministic to probabilistic \cite{Fung2008}. For example, random mutagenesis has recently shown to be succesful in developing metalloproteins \cite{Yang2018}. A range of machine learning techniques have been applied to site prediction in the past decade \cite{Akcapinar2017,Brylinski2011,Lin2006,Sodhi2004} although they are not prevalent within the design task outside of computationally validating designed sequences before experimental characterization.

Another task within the realm of protein design is designing entirely new protein topologies. It is challenging but opens up new applications as well as exploring the limits of the available fold space \cite{Huang2016,Kuhlman2003,Dagliyan2013,Taylor2009}. Similar to metalloproteins, the design process also has not yet seen pervasive use of machine learning algorithms.

Typical workflows for the design of novel metalloproteins, and novel protein topologies, consist of designing a complex template followed by computational selection of the best designs before experimental validation. For metalloproteins this also includes choosing a metal and identifying relevant binding residues. The computational selection typically relies on computationally costly force-field based molecular dynamics (MD) simulations \cite{Akcapinar2017,Huang2016}. This makes improving the efficiency of the computional pipeline a valuable pursuit.

Outside of protein design, another field that has seen recent development is that of deep learning research. An aspect of this development has been the rise of powerful new generative methods \cite{GAN,Kingma,Rezende,pixelRNN} that leverage deep architectures in order to learn complex distributions \cite{music,pixelCNN}. One of the most widely used deep generative methods, and the one used in this work, is the variational autoencoder (VAE). VAEs use a combination of an inference mechanism and a generative mechanism in order to learn a compressed and compact latent representation of the input data \cite{Kingma,Rezende,GomezBombarelli2018}. The generative mechanism can be used to sample complex synthetic data, and the inference mechanism allows sampling of synthetic data that is similar to a specified real input. Furthermore, this two mechanism structure allows for auxiliary tasks, like semi-supervised learning.


Deep generative models are only recently being applied to protein sequences \cite{Muller2018,Sinai2017} and this leaves a variety of potential applications not yet explored. For example, some work has been done using deep learning to generate protein sequences with desired features, however they do not use generative models. Instead they use deterministic deep models that take in a variety of engineered features to produce a sequence \cite{Li2014,Wang2018}. One example of using deep generative models to produce proteins has been published \cite{Muller2018}. This uses an autoregressive recurrent neural network trained on a database of antimicrobial peptides, however it does not take into account structure explicitly or learn a compact latent representation of the data.

In this work, a deep generative model is presented using conditional variational autoencoders (CVAE) for small single domain protein sequences. This model aims to reduce the initial search space by learning a distribution that narrows down the space of all possible proteins to a smaller space of proteins that are more likely to be natively ordered. This work explores two applications. Firstly, unsupervised design of metal binding sites in pre-existing proteins, and secondly producing proteins that fold into a novel topology.

For the metalloprotein design task the model is conditioned on metal binding ability for 8 different metals. Proteins are then designed by adding a new metal binding site where there had not been one. The second task takes inspiration from similar work which uses VAEs with data described by a grammar \cite{GrammarVAE}. Instead of producing novel small molecules like that study \cite{GrammarVAE}, the second task produces protein sequences. In order to condition the model in this task on a specified protein fold, a discrete representation of protein tertiary and secondary structure was required. This requirement was met by developing a context-free grammar (CFG) for protein fold structure. The grammar uses a `periodic table' of protein folds \cite{Taylor2002} from which a vectorized structural representation can be drawn. This task also uses both the generative and inference aspects of the VAE in an iterative method to explore the latent space, allowing sampling of sequences demonstrating a particular desired attribute.

Code and documentation required to reproduce the results of the paper and generate further sequences, along with a copy of the trained model, is freely available at \url{https://github.com/psipred/protein-vae}.

\section*{Methods}

\subsection*{Data collection}

Proteins are extracted from the Protein Data Bank (PDB) \cite{Berman2000} with the following criteria: monomers only, chain length no greater than 120 residues and no DNA/RNA present.
Only chains with single domains are retained by comparison with CATH \cite{Sillitoe2015}.
This gives 3,785 protein chains.
Homologues were found for each chain by running blastp \cite{Altschul1997} against the UniRef90 dataset of available sequences clustered at the 90\% similarity level \cite{UniProtConsortium2017}.
An E-value threshold of $10^{-3}$ was used and up to 50 hits were added to the dataset for each PDB structure.
The length of blastp hits was limited to 80\%-120\% of the query sequence with an upper limit of 140 residues.

Information on metal binding is collected from MetalPDB \cite{Andreini2013} for 8 metals - Fe (bound to 10\% of proteins in dataset), Zn (6\%), Ca (4\%), Na (2\%), Cu (2\%), Mg (1\%), Cd (0.9\%) and Ni (0.5\%).
For the model conditioned on metal binding sites, another sequence is added to the dataset to act as a negative training example for each sequence in the dataset that binds a metal.
This sequence is identical except that metal binding residues identified in MetalPDB are mutated to a random amino acid drawn from a distribution where each amino acid has the same frequency as in the whole training set.
For homologous sequences the residues in the corresponding places in the multiple sequence alignment are mutated if they are the same amino acid as in the PDB structure.
The dataset used for the metal binding model has 148,000 entries including the homologous sequences and negative training examples.

To construct the hidden Markov model (HMM) used for comparison to the VAE, hmmbuild from HMMER \cite{Eddy2011} was run with default parameters on a multiple sequence alignment (MSA) generated from the results of a blastp query of the sequence.

\subsection*{Context free grammar}

\begin{figure}
    \centering
    \includegraphics[width=0.90\textwidth]{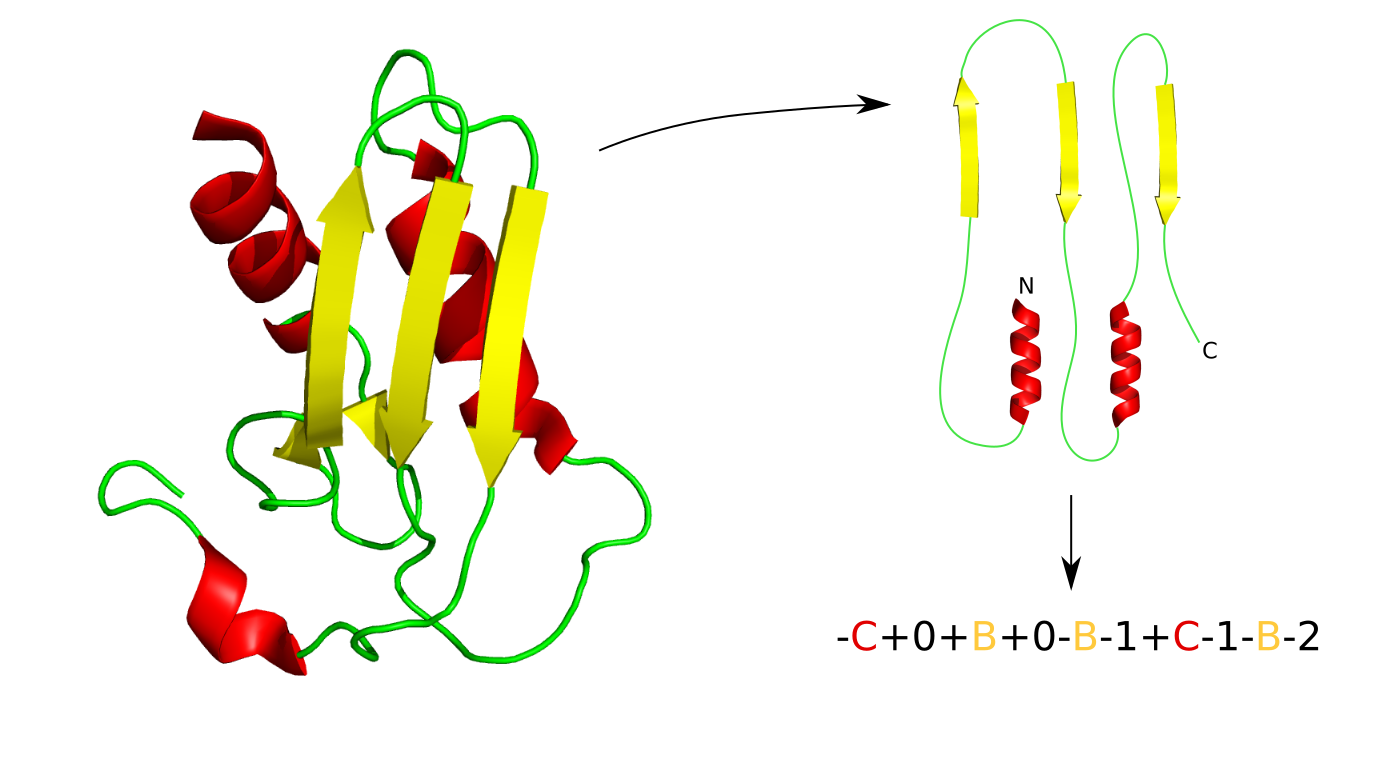}
    \caption{The Taylor grammar of protein structures shown for a reductase-related bacterial protein (PDB ID 2CU6).
    The orientation of the main secondary structural elements is examined in order to assign a topology string.
    See Taylor 2002 \cite{Taylor2002} for a full description of the grammar.}
    \label{fig:grammar}
\end{figure}

A grammar for protein structures allows conditioning of the output of a VAE.
We base our grammar on Taylor's `periodic table' of protein structures \cite{Taylor2002}.
This considers protein structures to belong to one of three basic forms (\textalpha \textbeta \textalpha\ layer, \textalpha \textbeta \textbeta \textalpha\ layer and \textalpha \textbeta\ barrel) with secondary structural elements added or removed to form individual topologies.
The formal context-free grammar that describes the topology strings is expressed as $G = \{ N, \Sigma , P, S \}$ where:
\begin{itemize}[noitemsep]
    \item Non-terminal symbols, $N = \{$
    \begin{itemize}[noitemsep]
        \item `\textless topology\textgreater ',
        \item `\textless element\textgreater ',
        \item `\textless orientation\textgreater ',
        \item `\textless layer\textgreater ',
        \item `\textless position\textgreater '
    \end{itemize}
    $\}$
    \item Terminal symbols, $\Sigma = \{$
    \begin{itemize}[noitemsep]
    	\item `+', `-',
        \item `A', `B', `C', `D',
        \item `+0', `+1', `+2', `+3', `+4', `+5', `+6',
        \item `-1', `-2', `-3', `-4', `-5', `-6'
    \end{itemize}
    $\}$
    \item Production rules, $P = \{$
    \begin{itemize}[noitemsep]
        \item `\textless topology\textgreater ' $\rightarrow$ `\textless element\textgreater \textless topology\textgreater ',
        \item `\textless topology\textgreater ' $\rightarrow$ `\textless element\textgreater ',
        \item `\textless element\textgreater ' $\rightarrow$ `\textless orientation\textgreater \textless layer\textgreater \textless position\textgreater ',
        \item `\textless orientation\textgreater ' $\rightarrow$ [`+', `-'] (2 rules),
        \item `\textless layer\textgreater ' $\rightarrow$ [`A', `B', `C', `D'] (4 rules),
        \item `\textless position\textgreater ' $\rightarrow$ [`+0', `+1', `+2', `+3', `+4', `+5', `+6', `-1', `-2', `-3', `-4', `-5', `-6'] (13 rules)
    \end{itemize}
    $\}$
    \item Start symbol, $S =$ `\textless topology\textgreater '
\end{itemize}
For example, the reductase-related bacterial protein with PDB ID 2CU6 fits form 0-3-2 (\textalpha \textbeta \textalpha) with topology string \mbox{`-C+0+B+0-B-1+C-1-B-2'}, where B and C are the layers prefixed by their relative orientation to the first strand in the sheet and suffixed by their position relative to the first element in each layer \cite{Taylor2002}.
This example is shown in Figure~\ref{fig:grammar}.

Proteins in our dataset were assigned to topology strings using an assignment of topology strings to SCOP folds \cite{Murzin1995, Taylor2017}.
Each protein was compared to all assigned representative SCOP proteins and the highest TMAlign score above 0.6, indicating a high chance of the same fold, was used to select the topology string to assign.
This allowed assignment for 65\% of proteins in the dataset described above to 325 unique topology strings.
This 65\%, equating to 105,000 sequences including homologues, was used to train the CVAE for the fold generation model.

\subsection*{Model}
\begin{figure}
    \centering
    \includegraphics[width=0.90\textwidth]{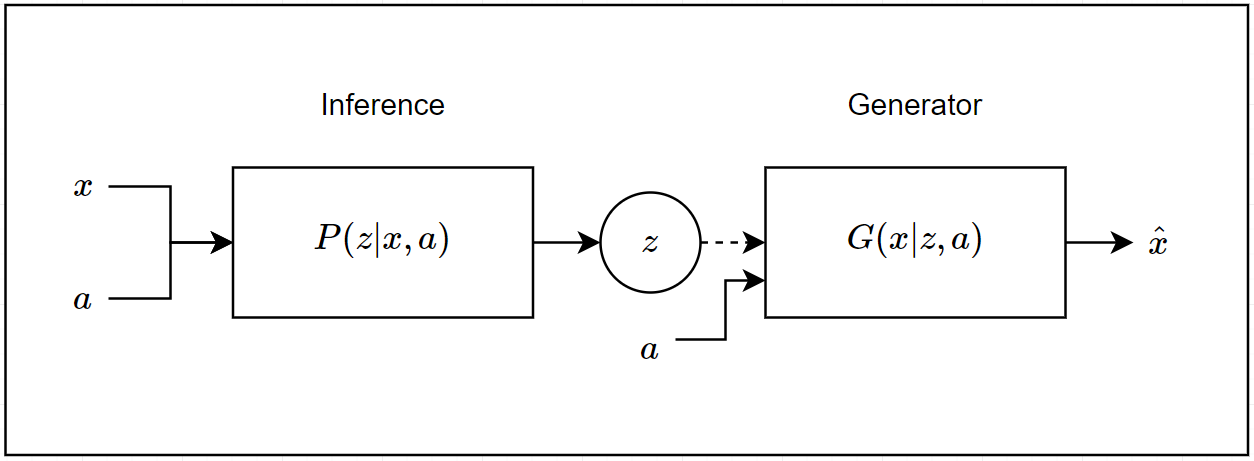}
    \caption{The inference/generator (encoder/decoder) structure of the VAE. The data $x$ and the conditioned attribute $a$ are concatenated and passed through the inference model to produce the latent code $z$. The attribute is then concatenated to the sampled $z$ (denoted by a dashed line) which going through the generator produces the reconstructed sequence $\hat{x}$.}
    \label{fig:vae_model}
\end{figure}

In this section we describe the how a variational autoencoder can be used to produce novel protein sequences when conditioned on an attribute like a grammar or a metal code.
VAEs simultaneously train two models, an inference model $q_\phi(z|x)$ and a generative model $p_\theta(x|z)p_\theta(z)$ for data $x$ and the latent variable $z$. For this application both models are also conditioned on a chosen attribute of the sequences, $a$ - see Figure~\ref{fig:vae_model}. Both models are jointly optimized using the tractable variational Bayes approach which maximizes the evidence lower bound (ELBO). 
\begin{align}
\log{p(x|a)} &\geq E_{q_\phi (z|x,a)} \Big[ \log \frac{p_{\theta}(x|z,a)}{q_\phi (z|x,a)}\Big] = \mathcal{L}(\theta, \phi; x) \\
&=E_{q_\phi (z|x,a)} \big[ \log p_{\theta}(x|z,a) \big] -KL(q_\phi (z|x,a)||p_\theta (z))
\end{align}
This equates to minimizing the reconstruction loss on $x$ and the Kullback-Leibler (KL) divergence between the inference model and a prior $p(z)$ usually characterized by an exponential family distribution, most typically a standard Gaussian.  
\begin{align}
 q_\phi(z|x,a)&=\mathcal{N}(z|\mu_q(x,a),\sigma^2_q(x,a))\\
 p_\theta(z)&=\mathcal{N}(z|0,\mathbb{I})
\end{align}
Using the reparameterization trick \cite{Kingma,Rezende}, which incorporates the prior using a linear transform of noise, ensures that both models are differentiable and can be optimized using stochastic gradient descent (SGD).

In terms of layerwise construction of the model we define a linear block (LB) as the following
$$
\texttt{LB}(x) = \texttt{ReLU}(\texttt{BN}(Wx+b))
$$
Where ReLU is a rectified linear unit and BN is batch normalization \cite{batchNorm}. It was found that inclusion of batch normalization improved reconstruction accuracy, which has been documented particularly in the case of VAEs \cite{LVAE}. 

We further define a multi-layer perceptron (MLP) as three sequential LBs. Both the decoder and the encoder contain one MLP. In the case of the decoder, the MLP is followed by a linear layer with a sigmoid activation function that produces the output sequence. The MLP in the encoder produces the Gaussian parameters as follows, where it is assumed $a$ and $x$ are concatenated as $y$:
\begin{align}
\mu &= \texttt{Lin}(\texttt{MLP}(y)) \\
\sigma^2 &=\texttt{Softplus}(\texttt{Lin}(\texttt{MLP}(y)))
\end{align}
Where \texttt{Lin} is a linear layer. Different sizes of latent space were tried and the final chosen for sequence generation is 16 dimensions. The sizes of the layers within the MLP are 512, 256, \& 128 units. The order as presented is used for the encoder but reversed for the decoder. 

Each sequence is represented as a series of one-hot encoded tensors with 22 possible positions. This represents the 20 amino acids, a padding symbol to reach max length, and a symbol for any non-standard amino acid such as selenocysteine. Given $L_{max}=140$ the sequences are sized as $140 \times 22$ before being flattened to $3080$ at which point they are inserted into the network while being concatenated with a tensor representing the conditioning attribute. In the case of metal binding sites this is a 1D tensor containing 8 values. Each is set to either zero or one to denote the absence or presence respectively of one of the eight metal binding sites. For the structure grammar this is a 1265 long flattened tensor of the one-hot encoded rules. Each rule is described by a one-hot tensor with 23 options, corresponding to the 22 rules $P$ and a blank rule for padding, and up to 55 rules are able to produce any given topology in the dataset. When converting produced sequences to their amino acid representations tensors are resized from $3080$ to $140\times 22$ and then the $\argmax$ is taken across the second dimension to choose the most likely amino acid for each position, leaving a sequence $140$ long.  

The model was optimized using an Adam optimizer \cite{Adam} with a learning rate of $5\times 10^{-4}$, a batch size of 512, and $\beta$ values of $0.9$ and $0.9$. These values were optimized by grid search. The KL divergence was set to zero at the beginning of training before being multiplied by a linearly increasing ``burn-in'' factor until the full KL loss was incorporated. This is to avoid the documented behavior of the KL loss turning off layer units early in training \cite{LVAE}. 

Models were trained until convergence. Furthermore all models were built using the Pytorch \cite{pytorch} framework for the Python programming language. Training time for the final model took roughly 6 hours to converge using a NVIDIA GTX 1080 TI GPU.

\subsection*{Sampling procedure - structure task}
With a maximum sequence length of $140$ residues there are $20^{140}$ possible proteins implying that even a reduced sequence space for a particular structure will likely contain a very large number of potential sequences \cite{Tian2017}. It has also been shown that within the protein space the optimal set of sequences for a particular structure is surprisingly small \cite{Kuhlman2000}. This being the case a sequential sampling-analysis-sampling method was used to traverse the space towards finding the region containing sequences that belonged to this optimal set.

\begin{figure}
    \centering
    \includegraphics[width=0.80\textwidth]{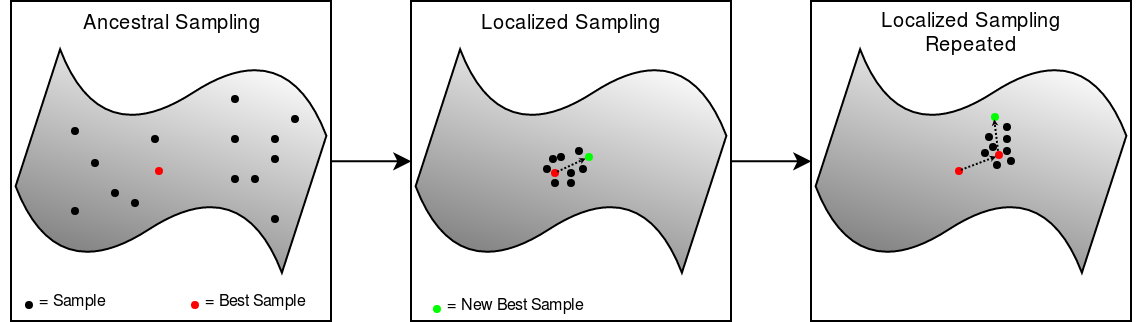}
    \caption{Graphic showing how the latent space was explored through iterative sampling and analysis of protein sequences.}
    \label{fig:sampling}
\end{figure}

For designing a new topology, the generator was sampled to produce 1,000 different sequences from across the space. From there sequences were analyzed, as described below, to find the best sequence of those generated. This sequence is then fed into the inference mechanism to find the mean and standard deviation tensors that define its latent code. From these are then used to sample 1,000 more samples using the generative network. This is analogous to searching the space around the best sequence to find potentially better sequences in the local area. This process is repeated several times to traverse the search space before finding a final best sequence - see Figure~\ref{fig:sampling}. 

In order to select promising sequences at each iteration, generated sequences are threaded onto a relevant template structure for the desired fold using SCWRL4 \cite{Krivov2009} with default options.
Energy minimisation is carried out using the Rosetta relax protocol \cite{LeaverFay2011} with the thorough option and 3 structures generated per sequence.
The top 5 sequences by score are taken forward for ab initio structure generation using Rosetta.
Fragment generation is carried out and 10,000 structures are generated using the AbinitioRelax protocol with default parameters.
Structures close to the template are found using TMAlign and the best sequence is input to the VAE in a new cycle.
Each iteration step takes around 500 hours of computation on a single CPU but is easily parallelized, making the iteration process computationally tractable.

\subsection*{Sampling procedure - metalloprotein task}

For design of metalloproteins the sampling process is simpler as removing or adding a binding site is done with a preexisting sequence. The protein chosen is fed into the inference mechanism with its conditioning attribute and 1,000 samples are then drawn from the generator using its mean and standard deviation tensors. 

To select the best candidate sequence, while limiting human oversight, a neural network classifier was trained to predict metal binding potential using the same dataset employed in training the VAE. Instead of having the metal binding code $c$ as the conditioning set it was used as a target set in a supervised fashion i.e. predicting $c$ given sequence $x$. 

The classifier was built using the same linear blocks used for the VAE (six sized 1024, 512, 256, 128, 64 \& 8 hidden units respectively) where the last layer utilized a sigmoid activation function instead of a ReLU function. This is because predicting metal binding is a multi-class classification task, each class being the protein's ability to bind one of eight metals. The loss function used was binary cross entropy and optimization was carried out using the same parameters as the VAE. The inverse ratio of the number of proteins that bind a given metal was used to weight the loss function to reduce the effect of imbalanced classes. 

Cross-validation was performed by creating a $90\%$/$10\%$ split between the training set and a validation set. To prevent proteins from the same families existing in both sets ECOD \cite{Cheng2014} was used. This was done by making sure no protein sequence in the validation set was a member of the same ECOD family as any protein in the training set. To prevent overfitting during training early stoppage was performed by saving after every epoch and choosing the model that best minimized the validation loss.
The MDM2 and TSG-6 sequences explored in the results were also placed in the validation set to avoid them being used for training, though only non-metal binding sequences are present for these proteins. The trained classifier was given the 1,000 sampled sequences and the sequence with the highest predicted metal binding potential for the selected metal was chosen as the final candidate sequence.

\subsection*{Molecular dynamics}

All molecular dynamics (MD) runs were carried out using the GROMACS package \cite{Abraham2015}.
Energy minimisation was conducted using a steepest descent energy minimisation of 5,000 steps in a vacuum and the OPLS-AA force field.
MD runs were conducted using periodic boundary conditions, SPC water, charge-neutralising counter ions, the OPLS-AA force field and a 2 fs timestep.
An initial energy minimisation was followed by a constant temperature and volume equilibration for 100 ps, then a constant pressure and temperature equilibration for 100 ps.
Production MD was run for 200 ns.

\section*{Results}

\begin{figure}
    \centering
    \includegraphics[width=0.90\textwidth]{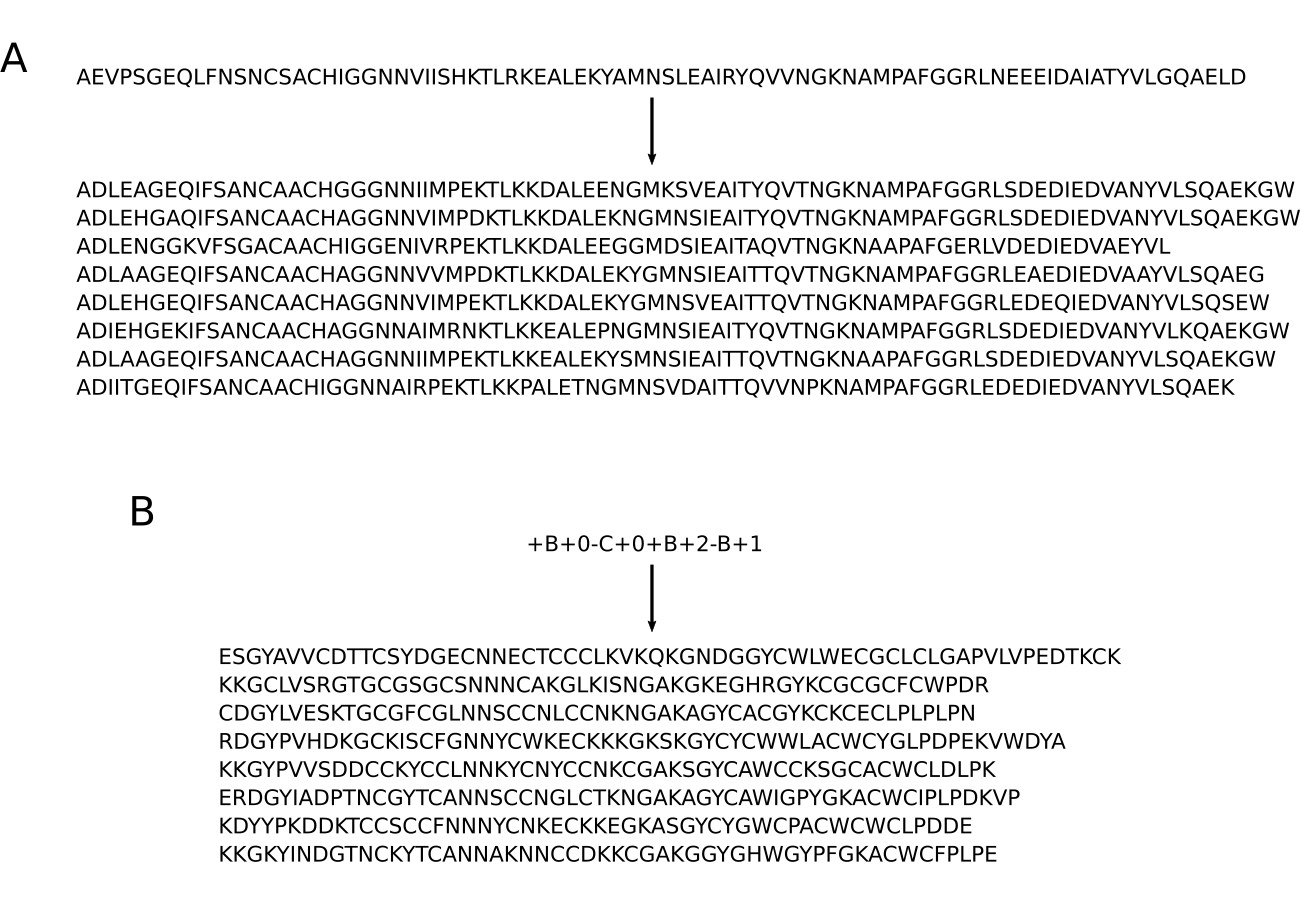}
    \caption{Example protein sequences generated by the model.
    (A) A sequence can be given as input, in which case similar sequences are returned by encoding and decoding the input sequence.
    (B) A topology string can be given as input, in which case sequences are generated that the VAE believes match the topology.}
    \label{fig:example_seqs}
\end{figure}

In order to carry out protein design tasks we developed a CVAE that is able to generate protein sequences with certain properties.
At its simplest the model is able to act as an encoder and decoder of protein sequences, with a mean sequence identity between training set sequences and their encoded-decoded form of 49.5\% for the model conditioned on metal binding sites.
Example sequences generated by the model are shown in Figure~\ref{fig:example_seqs}.
The conservation of residues across protein families is broadly reproduced by the model.
Residues conserved by evolution are generally not varied by the model, whereas residues not conserved by evolution are varied in the sequence output.
For example, a MSA was formed from 1,000 sequences generated from one lysozyme sequence (PDB ID 1IIZ) in the dataset using the model conditioned on the grammar.
The conservation per residue as given by Jensen-Shannon divergence \cite{Capra2007} shows a Pearson correlation coefficient of 0.72 with values from a MSA from a blastp query of the sequence.

\subsection*{Generation of metal binding sites}

The model is able to add potential metal binding sites to proteins.
The SWIB domain of human MDM2, a protein in the dataset used to train the model, was investigated as it is not known to bind metal ions and it has high sequence identity when encoded and decoded by the model.
1,000 sequences were generated by encoding and decoding this sequence using the VAE with the copper flag turned on, i.e.\ copper binding was requested.
In order to explore the benefits of using a VAE to generate sequences, we compared these sequences to those generated by a HMM.
A HMM produces sequences from an MSA assuming a Markov process with unobserved states.
Sequences generated with the VAE using the MDM2 sequence as input show less variability than sequences generated with a HMM produced from a MSA of the blastp hits of MDM2.
This is expected as we are using a single sequence as input rather than a protein family.

In order to assess the stability of generated sequences in the structure of the input sequence, sequences were threaded through this structure (PDB ID 3LBL) and the Rosetta energy scores after relaxation were compared.
As we are comparing different sequences adopting the same structure, a lower Rosetta energy score indicates a better suitability to the given structure.
The median score is -182 energy units for the VAE sequences, -115 for the HMM sequences and -184 for the native sequence.
This indicates that sequences generated by the VAE are likely to fold to the same structure as the input sequence, whereas the HMM sequences are likely to adopt a different structure or not fold to a stable structure.
Hence HMM sequences show more sequence and structural variation than VAE sequences.

When a copper binding site is requested in the VAE, more copper-binding motifs \cite{Regan1995} are observed in the output sequences than for the HMM.
10\% of sequences have histidine residues close in space (8\% for the HMM) and 4\% of sequences have His-x\textsubscript{3}-His on a \textalpha -helix (2\% for the HMM).
When the output sequences are limited to those with more histidines than the native sequence, 42\% of sequences from the VAE have close histidines (17\% for the HMM) and 16\% of sequences have His-x\textsubscript{3}-His on a \textalpha -helix (4\% for the HMM).
This indicates that the VAE is able to generate sequences with a high chance of folding to the same structure as the input sequence, yet can still add metal binding motifs to the sequences.
Whilst copper-binding motifs are not being explicitly requested from the HMM, fewer copper binding motifs are seen than for the VAE despite the higher sequence variability.
The ability to generate sequences with histidines close in space and with copper-binding motifs on helices shows that the VAE is learning structural information.

\begin{figure}
    \centering
    \includegraphics[width=0.99\textwidth]{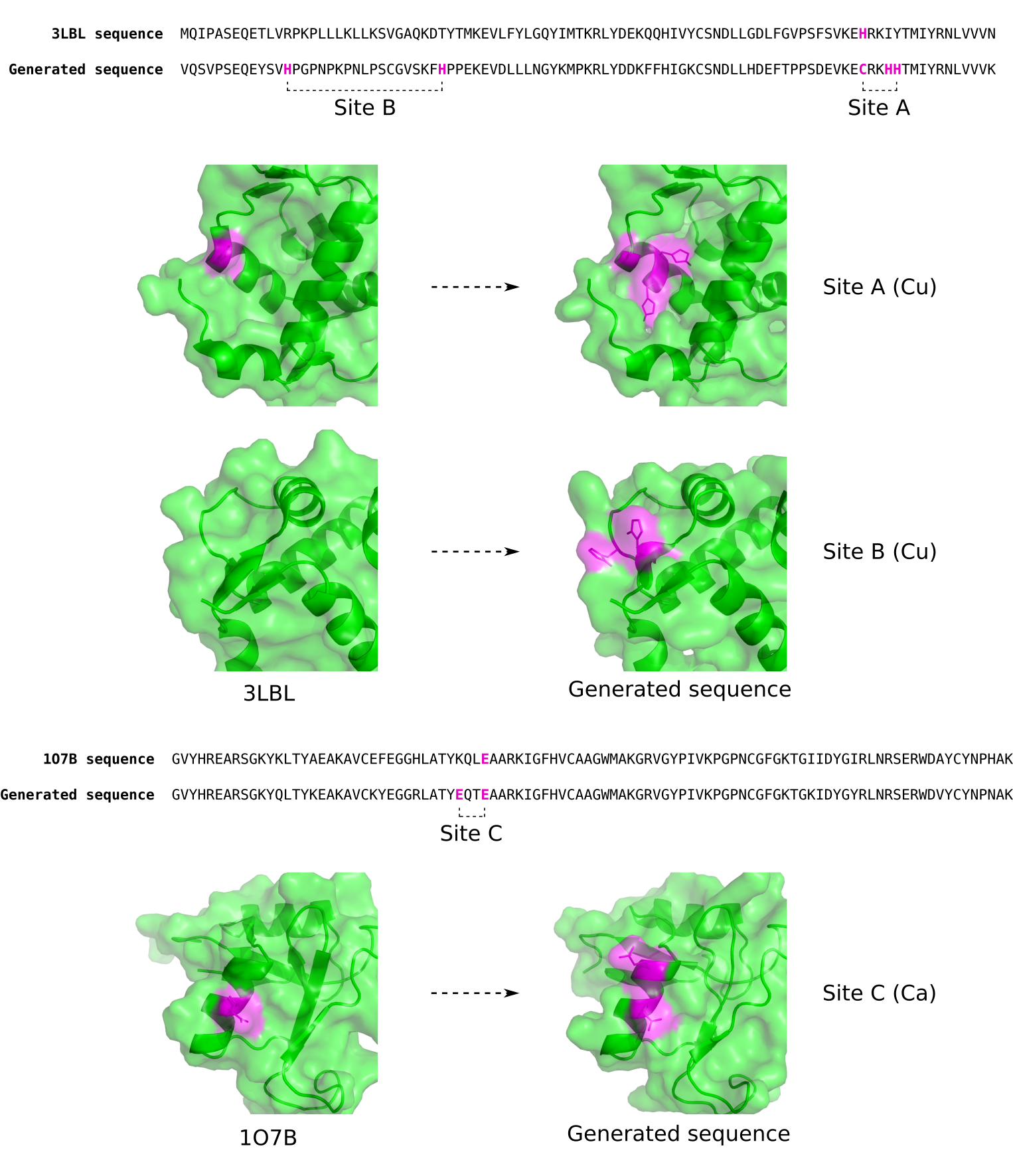}
    \caption{Automated addition of potential metal binding sites to two proteins.
    The sequence of the SWIB domain of human MDM2 (PDB ID 3LBL) and a sequence generated by the model that is predicted to have high copper-binding character are shown.
    Highlighted in purple are the residues that form the potential copper binding sites.
    The sites are shown on the structure of the generated sequence using 3LBL as a template, and compared to 3LBL.
    The same is shown for a potential site on the link module of human TSG-6 (PDB ID 1O7B) when calcium binding is requested.}
    \label{fig:metal_binding}
\end{figure}

A discriminator trained to predict metal binding sequences (see the Methods) was used to predict the sequence with the highest copper binding character from the 1,000 generated.
The top-ranked sequence has two potential new copper binding sites when compared to the input sequence, as shown in Figure~\ref{fig:metal_binding}.
One of the potential sites, site A, involves the modification of a site with a histidine residue to a site with two histidines and a cysteine to form a linear motif.
The other potential site, site B, involves the addition of two histidine residues close in structure but 18 residues apart in sequence to form a non-linear motif.
This sequence has 54\% identity to the input sequence.

Another protein, the link module of human TSG-6 (PDB ID 1O7B), was selected for similar reasons to the SWIB domain.
In this case calcium binding was requested.
The generated sequences show more aspartate residues, known to bind calcium \cite{Andreini2013}, than the native sequence in 29\% of cases.
The metal binding discriminator was used to select the sequence from 1,000 generated sequences most likely to bind calcium.
The second-highest ranked sequence shows a potential new calcium binding site formed of a glutamate added by the model and a native glutamate 3 residues apart, as shown in Figure~\ref{fig:metal_binding}.
This sequence has 89\% identity to the input sequence.
In both cases above a single sequence was used as input and a sequence was returned that shows potential metal binding character.
These results suggest that the model is able to add metal binding sites without having to manually examine individual protein structures.

Another test of the model's ability to add metal binding sites is whether it can add a known site back to a protein excluded from the training set.
This was tested with the classic copper-binding protein plastocyanin.
The instances of plastocyanin with native sequences were removed from the dataset; the instances with metal binding residues mutated to other amino acids were left in.
When copper binding sequences are requested with the mutated plastocyanin sequence as input from the model trained on this new dataset, 87\% of 1,000 sequences have the crucial copper-binding residue His37 present even though it was mutated in the training examples.
11\% of the sequences have the full copper-binding site of His37, Cys84 and His87 present.

\subsection*{Generation of proteins with a given topology}

Next, we explore the ability of a CVAE to generate sequences for a given protein topology.
In this case, the output is conditioned on a grammar of protein structures \cite{Taylor2002}.
Note that when generating sequences for a given topology only the topology is used as input; this is different to the metal binding case above when a protein sequence was used as input along with the metal binding information.
First, sequences were generated for a topology common in the dataset, the two-layer sandwich with three \textbeta -strands and one \textalpha -helix (topology \mbox{`+B+0-C+0+B+2-B+1'} - see the Methods).
In this case the sequences generated show similarity to the training sequences with the given topology, with 55\% of generated sequences having 50\% or greater sequence identity to at least one training sequence.
Threading generated sequences through four PDB structures with the given topology (PDB IDs 1AHO, 1JZA, 2FKL and 2M8B), followed by Rosetta energy minimisation, gives energy scores that are similar to the PDB structures in many cases.
18\% of generated sequences have energy scores within 10\% of the native PDB scores for at least one of the above four structures, which only represent a subset of available structures of the fold.
In this case the CVAE seems to be generating viable homologues of the sequences used for training.

\begin{figure}
    \centering
    \includegraphics[width=0.90\textwidth]{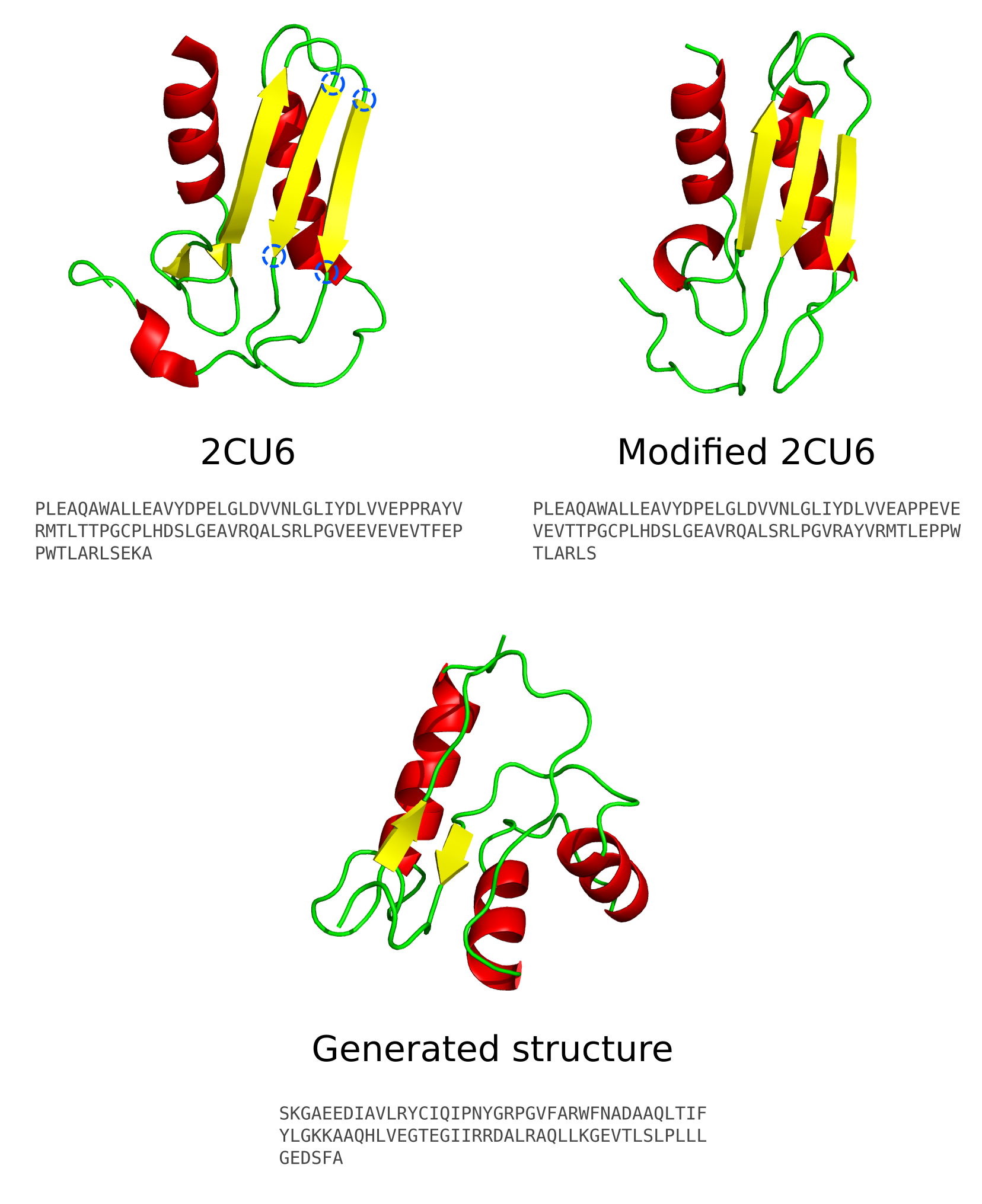}
    \caption{Structures to explore a novel fold.
    The structure and sequence of 2CU6 are shown.
    By remodelling the loops at the locations of the blue circles using ModLoop \cite{Fiser2003} a modified structure with a novel topology is generated.
    This is used as a backbone template to select structures from a pool of ab initio Rosetta structures generated from sequences output by the CVAE.
    The closest structure to the template is shown.}
    \label{fig:novel_fold}
\end{figure}

Next, the model was used to generate sequences for a novel topology.
This topology was made by picking a structure with a well-assigned Taylor topology - the reductase-related bacterial protein with PDB ID 2CU6 - and modifying the loops connecting the central \textbeta -strands with ModLoop \cite{Fiser2003}.
See Figure~\ref{fig:novel_fold}.
The secondary structural elements are arranged identically in 3D space but the different connection of the loops creates a new topology and leads to a large rearrangement of the sequence.
The resulting topology (\mbox{`-C+0+B+0-B-2+C-1-B-1'}) is not assigned to any structure in the PDB and a structural search with the modified protein using DALI \cite{Holm2010} does not give any close matches, the only hits being similar in topology to 2CU6.

Sequences were generated for this topology using the model, and an iterative procedure of selecting promising sequences via ab initio structure generation and localized sampling was used to refine the sequence generation (see the Methods).
After three iterations of this process a sequence of 86 residues was generated that had structures in the pool of ab initio predicted structures similar to the modified 2CU6 structure.
This is shown in Figure~\ref{fig:novel_fold}.
The region between the \textbeta -strand elements contains two \textalpha -helices rather than one and there is less \textbeta -strand character, but the overall topology of \textalpha -helices behind a \textbeta -strand with certain connections is correct.
In fact the PSIPRED \cite{Buchan2013} prediction of the region not forming a \textbeta -sheet in the generated structure is of a \textbeta -sheet, and with minor rearrangement of the structure a \textbeta -sheet could be formed.

Although the modified 2CU6 backbone was used to guide the search, no sequence information from 2CU6 was utilised and the sequence is generated purely by the CVAE.
There are no similar sequences in the PDB, searching with an E-value threshold of 10.
The sequence shows some similarity only to amidophosphoribosyltransferase sequences, with the highest scoring blastp hit having 46\% sequence identity over 55\% of the query protein. 
However structures of amidophosphoribosyltransferases in the PDB show no similarity to the requested topology.
Interestingly, the sequence shows 28\% sequence identity to a sequence generated by RosettaDesign \cite{Liu2006} from the generated structure backbone in Figure~\ref{fig:novel_fold}, indicating some agreement with established design protocols.

\begin{figure}
    \centering
    \includegraphics[width=0.99\textwidth]{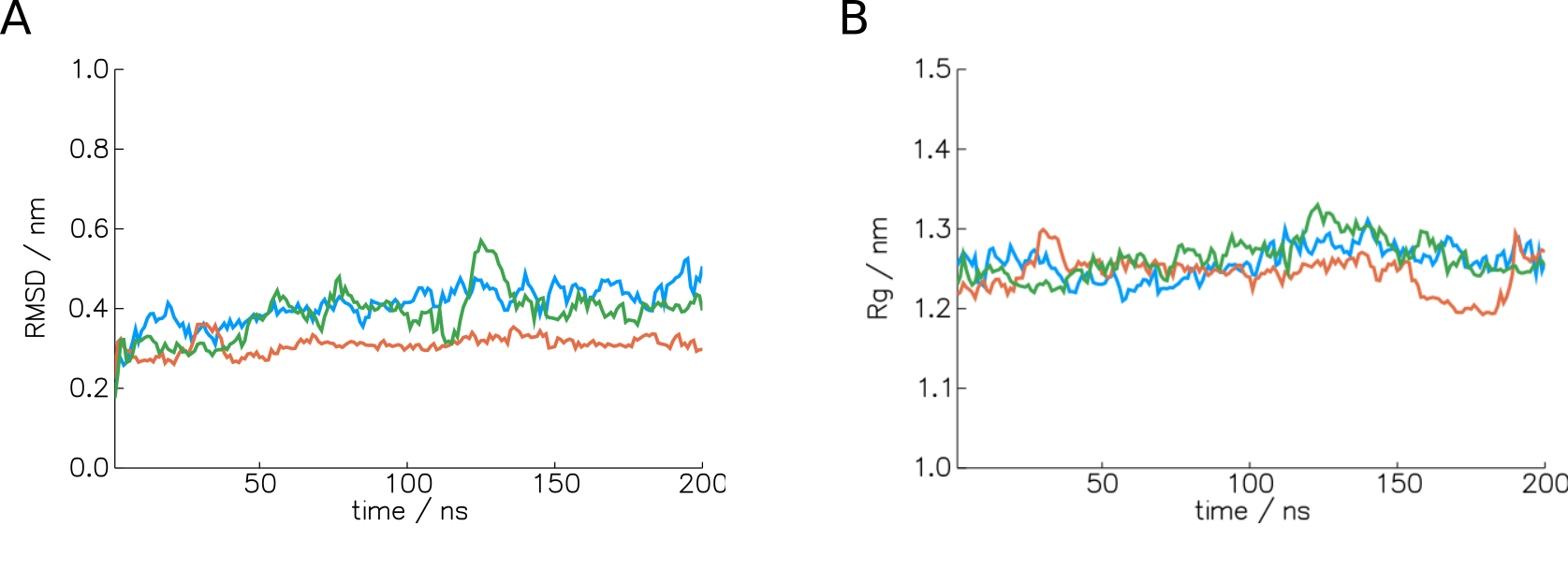}
    \caption{Analysis of MD runs of the generated structure shown in Figure~\ref{fig:novel_fold}.
    Three runs of 200 ns are shown in blue, orange and green.
    (A) Backbone RMSD of trajectory structures to the energy-minimized starting structure.
    (B) Backbone radius of gyration (Rg) of trajectory structures.}
    \label{fig:md_graphs}
\end{figure}

In order to probe the stability of this structure and gain information on whether the sequence adopts it, MD simulations were carried out.
MD simulations can lead to stabilisation of collapsed structures other than the native structure due to favouring hydrophobic interactions \cite{Robustelli2018}, but they are still a useful tool for probing potential structures \cite{Lindorff-Larsen2011}.
The structure showed little deviation in structure over three runs of 200 ns simulation time when considering root mean square displacement (RMSD) and radius of gyration.
This is shown in Figure~\ref{fig:md_graphs}.
Although longer simulations would be required to check more thoroughly for unfolding or for a different folded state, this is some indication that the fold is stable.
Experimental testing would be required to verify if this prediction is correct.
However, these results indicate that a CVAE is able to generate viable sequences for protein topologies both known and not previously seen.
It is hoped that these results generalize to the vast number of topologies described by the grammar.

\section*{Discussion}

\begin{figure}
    \centering
    \includegraphics[width=0.8\textwidth]{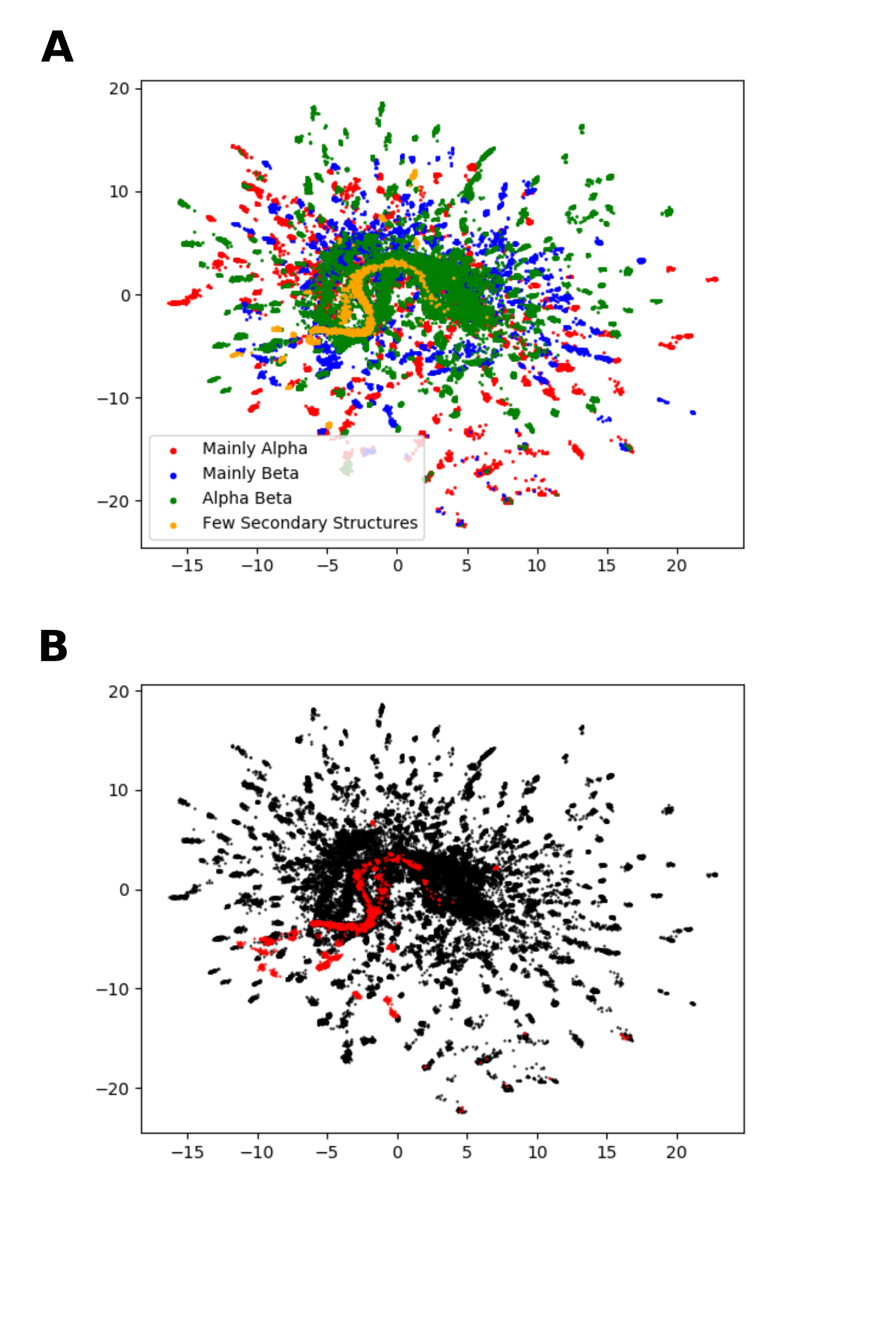}
    \caption{Separation by structural properties in the latent space when 2 latent dimensions are used in the model.
    The axes are the 2 latent dimensions and each point is the encoded representation in the 2 dimensions of one input sequence.
    Clusters generally correspond to the homologues collected for each sequence.
    (A) Each sequence is coloured by CATH class \cite{Sillitoe2015} according to the colours shown.
    (B) Sequences for one CATH architecture, `mainly beta single sheet' (CATH ID 2.20), are highlighted in red.}
    \label{fig:latent}
\end{figure}

This work has indicated that CVAEs are able to carry out protein design tasks by conditioning output sequences on desired properties such as metal binding sites and given topologies.
The model is able to learn structural properties, even with fewer latent dimensions than the 16 used in the final model.
For example, with 2 latent dimensions the homologues of each input sequence cluster together in the latent space.
The sequences also show some separation by CATH class, and sequences of the same CATH architecture tend to be close in the latent space.
These properties are shown in Figure~\ref{fig:latent}.
As would be expected, the first latent dimension correlates with molecular weight as the model learns the length of the input sequence.

Sequences generated by the model described here show similar conservation patterns to native sequence families.
However, the covariation of residues present across sequence families is not reproduced by the model.
This is to be expected for two reasons: each PDB sequence in the dataset is limited to 50 sequence homologues, and the input to the model is a single sequence rather than a multiple sequence alignment.
There are developments to the model that could lead to sequences being generated with realistic covariation signals. These include training directly on multiple sequence alignments, having a data set with proteins that have more homologues, and using regularization techniques to promote covarying outputs.
It has been shown that taking into account covariation is important for effective protein design \cite{Socolich2005, Tian2018}.
Another potential use of this is to generate additional sequences for use in residue contact prediction methods.

Our model has shown promise in conditioning output sequences in terms of metal binding or a given topology.
There are other properties of proteins that could be explored using the model in a similar way.
These include protein-protein interactions, which have been organised systematically in an analogous way to Taylor topologies \cite{Ahnert2015}; allosteric sites, which have been designed into proteins previously \cite{Dagliyan2013}; taking account of multiple conformations in the design process \cite{Davey2017, Ambroggio2006}; and designing proteins by linking together large fragments from the PDB \cite{Jacobs2016}.
Another approach is to use activation maximization \cite{Nguyen2016} on trained models to directly move towards sampling proteins that have measurable and desired attributes. For example, sampling a sequence containing a particular motif.    

After years of effort, protein design is becoming a systematic tool showing promise in many areas.
However, the difficulties of navigating a vast search space and scaling design efforts from specific proteins to the whole of protein space remain a challenge.
The rise of deep learning has the potential to meet this challenge due to its ability to learn complex patterns without supervision and the small computational cost of using a model once it is trained.
In addition, the increasing amount of sequence \cite{Kunin2008, Asgari2015} and structural \cite{Callaway2015} data will aid data-intensive methods such as deep learning.
It is hoped that methods similar to the CVAE presented here will be a valuable addition to the protein design pipeline.

\section*{Data availability}

Code and documentation required to reproduce the results of the paper and generate further sequences, along with a copy of the trained model, is freely available at \url{https://github.com/psipred/protein-vae}.

\section*{Acknowledgements}

We thank members of the group for valuable discussions and comments, and William R Taylor for providing data and assistance for the protein structure grammar.
This work was supported by the European Research Council Advanced Grant `ProCovar' (project ID 695558) and the Biotechnology \& Biological Sciences Research Council (BBSRC) UK, grant BB/M011712/1.
This work was supported by the Francis Crick Institute which receives its core funding from Cancer Research UK (FC001002), the UK Medical Research Council (FC001002), and the Wellcome Trust (FC001002).

\section*{Author contributions statement}

JGG, LM and DTJ conceived and designed the study and reviewed the manuscript.
JGG and LM carried out the computational work, did the analysis and drafted the manuscript.

\section*{Additional information}

\subsection*{Competing interests}

The authors declare no competing interests.

\end{document}